\documentclass[twocolumn,english,aps,prb]{revtex4}
\usepackage{bm}  
\usepackage{graphicx}
\usepackage{amssymb}
\usepackage{color}
\usepackage{amscd}
\usepackage{epsfig}

\newcommand{\bfr}{{\bf r}}

\newcommand{\nhat}{\hat{n}}
\newcommand{\nvec}{{\bf \hat{n}}}

\newcommand{\trho}{\tilde{\rho}}

\begin{document}
\title{Nematic and Polar order in Active Filament Solutions}
\author{A. Ahmadi$^1$}
\author{T. B. Liverpool$^2$}
\author{M. C. Marchetti$^1$}
\affiliation{$^1$Physics Department, Syracuse University,
Syracuse, NY 13244, USA} \affiliation{$^2$Department of Applied
Mathematics, University of Leeds, Woodhouse Lane, Leeds LS2 9JT,
UK}

\date{\today}

\begin{abstract}
  Using a microscopic model of interacting polar biofilaments and
  motor proteins, we characterize the phase diagram of both
  homogeneous and inhomogeneous states in terms of experimental
  parameters. The {\em polarity} of motor clusters is key in
  determining the organization of the filaments in homogeneous
  isotropic, polarized and nematic states, while motor-induced
  bundling yields spatially inhomogeneous
  structures.

\end{abstract}
\maketitle
\maketitle

Soft active systems are exciting examples of a new type of
condensed matter where stored energy is continuously transformed
into mechanical work at microscopic length scales. A realization
of this are polar filaments interacting with associated molecular
motors in the cell cytoskeleton \cite{Alberts,Howard}. These
systems are characterized by a variety of dynamic and stationary
states which the cell accesses as part of its cycle
\cite{takiguchi,nedelec97,surrey01}.

There have been a  number of recent theoretical studies of the
collective dynamics of isotropic and polarized solutions of active
filaments. These include numerical
simulations~\cite{nedelec97,surrey01}, mesoscopic mean-field
kinetic equations~\cite{nakazawa96,Kruse00,Kruse01,TBLMCM03}, and
hydrodynamic equations where the system is described in terms of a
few coarse-grained fields whose dynamics is inferred from symmetry
considerations~\cite{Lee01,ramaswamy02,Kruse04,Sankararaman04}.
Previous work has focused on how motor activity renders homogeneous
states unstable to the formation of spatial structures, such as
bundles, vortices or asters. In this article we study the profound
effect of motor activity on the possible homogeneous states of the
system \cite{Aranson05}. We also find important differences in the
nature of the instabilities from these homogeneous phases. Starting
from a microscopic model of interacting rigid filaments, we
characterize a phase diagram of homogeneous and inhomogeneous states
of active filaments in terms of experimentally variable parameters. We
find in particular that the formation of a non-equilibrium polarized
phase at low densities can be driven by motor {\em polarity} without
the need for filament polymerization~\cite{Lee01,Sankararaman04}.

\begin{figure}
\center \resizebox{0.40\textwidth}{!}{%
  \includegraphics{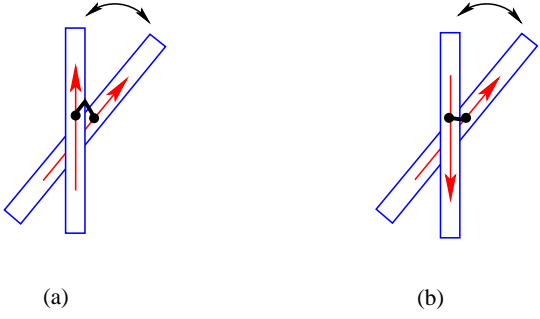}
}

\caption{ Polar and nonpolar clusters interacting with polar
  filaments. Assuming that clusters always bind to the smallest angle,
  polar clusters ($g \rightarrow \infty$) bind only to filaments in
  configuration (a) while non-polar clusters $(g=0)$ bind to both
  configurations equally.}
\label{interaction}       
\end{figure}

We describe the system by a concentration of polar filaments
$f(\bfr,\nvec,t)$ in two dimensions ($d=2$), modeled as hard rods
of {\em fixed} length $\ell$ and diameter $b$ ($\ell>>b$) at
position $\bfr$ with filament polarity characterized by a unit
vector $\nvec$, and a density of motor clusters $m(\bfr,t)$ at
position $\bfr$. The filament and motor concentrations satisfy the
equations
%
\begin{eqnarray}
\partial_tf &=&-\bm{\nabla}\cdot {\bf J}_f- \bm{\mathcal R} \cdot \bm{\mathcal J}\;, \\
\partial_t m &=& -\bm{\nabla} \cdot {\bf J}_m
\end{eqnarray}
where $\bm{\mathcal R} = \nvec \times \partial_\nvec$ and the
translational (${\bf J}_f,{\bf J}_m$) and rotational
($\bm{\mathcal J}$) currents have diffusive, excluded volume and
active contributions. The rotational current is
%
$\bm{\mathcal J}=\bm{\mathcal J}^{\rm D}+\bm{\mathcal J}^{\rm
X}+\bm{\mathcal J}^{\rm A}$,
%
with a diffusive current $\bm{\mathcal J}^{\rm
D}(\bfr,\nvec,t)=-D_r\bm{\mathcal R}f(\bfr,\nvec,t)$ and a
contribution from excluded volume, $\bm{\mathcal J}^{\rm
X}(\bfr,\nvec,t)=-\frac{D_r}{k_BT}f(\bfr,\nvec,t)\bm{\mathcal
R}V_{\rm X}(\bfr, \nvec,t)\;$, with
\begin{equation}
V_X(\bfr,\nvec_1,t)=k_BT\int_{s_1}\int_{s_2}\int_{\nvec_2}|\nvec_1\times\nvec_2|f(\bfr+\bm{\xi},\nvec_2,t)\;,
\end{equation}
where $\bm{\xi}=\nvec_1s_1-\nvec_2s_2$ is the separation of the
centers of mass of the two filaments and
$\int_s...=\int_{-\ell/2}^{+\ell/2}ds ...$ denotes an integration
along the length of the filament, parametrized by $s$. The active
contribution to the rotational current (low density approximation)
is
\begin{eqnarray}
\bm{\mathcal J}^{\rm A}(\bfr,\nvec_1,t)&=&b^2 \int_{s_1}
\int_{s_2} \int_ {\nvec_2}\bm{\omega}^A(\nvec_1,\nvec_2)~m(\bfr+
s_1 \nvec_1,t) \nonumber \\ && \times f(\bfr,\nvec_1,t)f(\bfr +
\bm{\xi} ,\nvec_2,t)\;,
\end{eqnarray}
where the motor induced angular velocity is written as
\begin{equation}
\bm{\omega}^A=2\Big(\gamma_0+\gamma_1\nvec_1\cdot\nvec_2\Big)(\nvec_1\times\nvec_2)
 \;. \end{equation}
It consists of two parts, corresponding to two classes of motor
clusters (see Fig.~\ref{interaction}): polar clusters which tend to
bind to filaments with similar polarity ( $\gamma_0/\gamma_1 \gg
1$)~\cite{nedelec97,surrey01}, and non-polar clusters which bind to
filament pairs of any orientation ( $\gamma_0/\gamma_1 \ll 1$)~\cite{kas02}.
Earlier work by two of us~\cite{TBLMCM03} considered only non-polar
clusters ($\gamma_0=0$)~\cite{gammaest}.  Both $\gamma_0$ and
$\gamma_1$ will increase with increasing binding rate of the clusters
to the filament. A passive polar crosslink (e.g. $\alpha$-actinin on
F-actin) will also have this effect~\cite{sackmann96}.
%
%
Since the binding rate can have different dependence on
ATP-consumption than the motor stepping speed, we
expect a different dependence on ATP hydrolysis rate than the active
contributions to the translational currents.

The translational currents are ${\bf J}_f={\bf J}^D+{\bf J}^X+{\bf
J}^A$ and
\begin{eqnarray}
{\bf J}_m= - D_m \bm\nabla m + \chi \int_s\int_{\nvec}\nvec
f(\bfr,\nvec,t)m(\bfr+\nvec s,t)\;,
\end{eqnarray}
where $\chi$ depends on  the speed and processivity of the motors.
 The filament
diffusive current, $J_{i}^D=-D_{ij}\partial_j \, f$, is expressed
in terms of a diffusion tensor
$D_{ij}=(D_\parallel+D_\perp)\delta_{ij}/2+(D_\parallel-D_\perp)\hat{Q}_{ij}$ with $\hat{Q}_{ij}=\nhat_i\nhat_j-\frac{1}{2}\delta_{ij}$.
The excluded volume contribution is $J^X_i=-\frac{D_{ij}}{k_BT} \,
f
\partial_jV_X$. The active contribution to the
translational current is
\begin{eqnarray}
J^A_i(\bfr,\nvec_1,t)&=&b^2\int_{s_1}\int_{s_2}\int_{\nvec_2}{
v}_i^A(\nvec_1,\nvec_2,\bm\xi)
m(\bfr+\nvec_1s_1,t)\nonumber\\
&&\times f(\bfr,\nvec_1,t)f(\bfr+\bm{\xi},\nvec_2,t)\;,
\end{eqnarray}
with ${\bf v}^A$  the motor-induced  velocity, taken of the form
\begin{equation}
{\bf
v}^A=\frac{\beta}{2}(\nvec_2-\nvec_1)+\frac{\alpha}{2}\frac{\bm\xi}{2
\ell}-\lambda(\nvec_1+\nvec_2)\;.
\end{equation}
The parameters $\alpha$, $\beta$ and $\lambda$ have dimensions of
velocity and depend on the angle between the filaments. The term
proportional to $\beta$ drives the separation of filaments of opposite
polarity, while the $\lambda$ contribution arises from the net
velocity of the filament pair \cite{meanVnote}. The negative sign
reflects the fact that filament mean motion due to motor activity is
opposite to their polarity. The contribution proportional to $\alpha$
arises from spatial variations in motor activity along the filament,
such as motors stalling before detaching at the polar end.  It drives
bundling of filaments of the same polarity.  These parameters where
estimated in Ref.  \cite{TBLMCMepl} via a microscopic model of
motor-induced filament dynamics as $\beta\sim\lambda\sim u_0$,
$\alpha\sim u_0(l_m/l)<<u_0$, with $u_0$ the mean motor stepping rate
and $l_m$ the length scale (of order of the motor cluster size) for
spatial variations in motor activity. As seen below, this term is
crucial for developing inhomogeneities and pattern formation.

To study the macroscopic properties of the solution, we truncate
the exact moment expansion of $f(\bfr,\nvec,t)$ as
\begin{equation}
f(\bfr,\nvec,t)=\frac{\rho(\bfr,t)}{2 \pi}\Big\{1+2 {\bf
p}(\bfr,t)\cdot\nvec+4S_{ij}(\bfr,t)\hat{Q}_{ij}\Big\}\;,
\label{tens_expand}\end{equation}
keeping only the first three moments, \begin{eqnarray}
&&\int d\nvec ~f(\bfr,\nvec,t)=\rho(\bfr,t)\;\; \mbox{(density)}, \nonumber\\
&&\int d\nvec ~\nvec~f(\bfr,\nvec,t)=\rho(\bfr,t){\bf p}(\bfr,t)\;\;\mbox{(polarization)}, \\
&&\int d\nvec
~\hat{Q}_{ij}~f(\bfr,\nvec,t)=\rho(\bfr,t)S_{ij}(\bfr,t)\;\;\mbox{(nematic
order)} \, . \nonumber \label{orient_moments}\end{eqnarray}

\paragraph{Homogeneous Bulk Steady States.}

We first consider the dynamical equations for a spatially
homogeneous solution. In this case the only contributions to the
equation of motion of the filament density come from rotational
currents. The motor density has a constant mean value (we let
$m_0=mb^2$) and the filament density, $f(\bfr,\nvec,t)$, and its
moments can be expressed in terms of their spatial averages, i.e.
${1 \over A} \int d\bfr S_{ij}(\bfr,t)=S_{ij}(t), {1 \over A} \int
d\bfr {\bf p} (\bfr,t)={\bf
  p}(t) $, with $A$ the area of the system.  In the following all lengths are measured in units of the
filament length, $\ell$.  Averaging over the orientation $\nvec$
using Eq.~(\ref{tens_expand})
, we find that in a homogeneous system $\rho=\rho_0={\rm constant}$
and
%
\begin{eqnarray}
\partial_tp_i&=&-\Big(D_r-m_0\rho_0\gamma_0\Big)p_i \nonumber \\
&&+\Big[\frac{8D_r
}{3\pi}-m_0\big(2\gamma_0-\gamma_1\big)\Big]\rho_0
S_{ij}p_j\;,\label{hydrop}\\
\partial_t
S_{ij}&=&-\Big[4D_r-\frac{8D_r\rho_0
}{3\pi}-m_0\rho_0\gamma_1\Big] S_{ij} \nonumber \\
&&+2m_0\rho_0\gamma_0\Big(p_ip_j-\frac{1}{2}\delta_{ij}p^2\Big)\;.
\label{hydroS}
\end{eqnarray}
%
In a passive system ($\gamma_0=\gamma_1=0$) there is a transition
from an isotropic state to a nematic state. A mean-field
description of such a transition, which is continuous in 2d (but
first order in 3d), requires that one incorporate cubic terms in
the nematic order parameter in the equation of motion. The
transition here is identified with the change in sign of the decay
rate of $S_{ij}$, which signals an instability of the isotropic
homogeneous state. This occurs when excluded volume effects
dominate at a density  $\rho_N= 3\pi/2$. The homogeneous state is
isotropic for $\rho_0<\rho_N$ and nematic for $\rho_0>\rho_N$. No
homogeneous polarized state with a nonzero mean value of ${\bf p}$
is obtained in a passive solution.

We now turn to an active system. We introduce a dimensionless
filament density, $\trho=\rho_0/\rho_N$, a dimensionless motor
cluster activity, $\mu=\rho_N m_0\gamma_1/D_r$, and a
parameter measuring the polarity of motor clusters, $g
=\gamma_0/\gamma_1$ with $g=0$ corresponding to non-polar
clusters. Time is measured in units of $D_r^{-1}$. The steady
states of Eqs.~(\ref{hydrop}) and (\ref{hydroS}) are the stable
solutions of
\begin{eqnarray}
0 &&=-a_1p_i
+b_1\trho S_{ij}p_j\;,\label{psteady} \\
0 &&=-a_2S_{ij} +b_2\trho
\Big(p_ip_j-\frac{1}{2}\delta_{ij}p^2\Big)\;, \label{Ssteady}
\end{eqnarray}
where
\begin{eqnarray}
&& a_1=1-\trho g\mu\;,\\
&& a_2= 4\left[1-\trho\left(1+ \mu  /4\right)\right]\;,
\end{eqnarray}
and $b_1=4+ \mu\left( 1-2g\right)$, $b_2= 2 g\mu$. At low density
the only solution is $p_i=0$ and $S_{ij}=0$ and the system is
isotropic (I). The homogeneous isotropic state can become unstable
in two ways. As in the passive case, a change in sign of the
coefficient $a_2$ signals the transition to a nematic (N) state.
Motor activity lowers the density for the I-N transition which
occurs at $\rho_{IN}(\mu)=1/(1+\mu /4)$.  At
$\trho>\rho_{IN}(\mu)$ the solution acquires nematic order, with
$S_{ij}^0=S_0(n_in_j-\delta_{ij}/2)$, where the unit vector ${\bf
n}$ denotes the direction of broken symmetry.
We obtain an expression for the amount of
nematic order ($S_0$) by adding a cubic term $-c_2 \tilde{\rho}^2 S_{kl}S_{kl}S_{ij}$ to  Eq.~(\ref{Ssteady}) giving $S_0 = {1 \over \tilde{\rho}}\sqrt{-2a_2/c_2}$. The isotropic
state can also become linearly unstable via the growth of
polarization fluctuations in any arbitrary direction. This occurs
above a second critical filament density,
$\rho_{IP}(\mu)=1/(g\mu)$, defined by the change in sign of the
coefficient $a_1$ controlling the decay of polarization
fluctuations. For $\trho>\rho_{IP}(\mu)$ the homogeneous state is
polarized (P), with $p_i\not=0$. The alignment tensor also have a
nonzero mean value in the polarized state as it is slaved to the
polarization. One can identify two scenarios depending on the
value of $g$.

\begin{figure}
\center \resizebox{0.40\textwidth}{!}{%
  \includegraphics{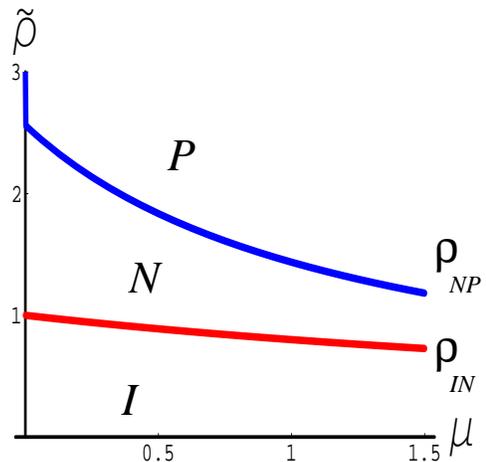} }

\caption{ The homogeneous phase diagram for $g<1/4$ (the figure is
for $g=1/10$ and $c_2=50$). The vertical axis is
  $\rho_0/\rho_N$, the horizontal axis is $\mu$. }
\label{PD1}       
\end{figure}

I) For $g < 1/4$, the density $\rho_{IP}$ is always larger than
$\rho_{IN}$   and a region of nematic phase exists for all values
of $\mu$. At sufficiently high filament and motor densities, the
nematic state also becomes unstable.
Fluctuations in the alignment tensor are
uniformly stable for $a_2<0$, but polarization fluctuations along
the direction of broken symmetry become unstable for $a_1\leq\trho
b_1 S_0/2$, i.e., above a  critical density
\begin{equation}
\rho_{NP}=\frac{1}{g\mu} \left[ 1 + {b_1^2 \over c_2 R}\left(  1 - \sqrt{1 + {2 c_2 R (1-R)\over b_1^2}}\right)\right]
\end{equation}
where $R = \rho_{IN} / \rho_{IP}$.
The polarized state at $\trho>\rho_{NP}$ has $p_i^0=p_0n_i$ and
$S_{ij}^0=S_P(n_in_j-\delta_{ij}/2)$, where
$p_0^2=\frac{2a_1a_2}{\trho^2
  b_1b_2}\big[1-\big(\frac{2a_1}{b_1S_0}\big)^2\big]$ and
$S_P=S_0\sqrt{1-\frac{\trho^2 b_1b_2}{2a_1a_2}p_0^2}=2|a_1/b_1|$.
The  "phase diagram" is shown in Fig.~\ref{PD1}.

\begin{figure}
\center \resizebox{0.40\textwidth}{!}{%
  \includegraphics{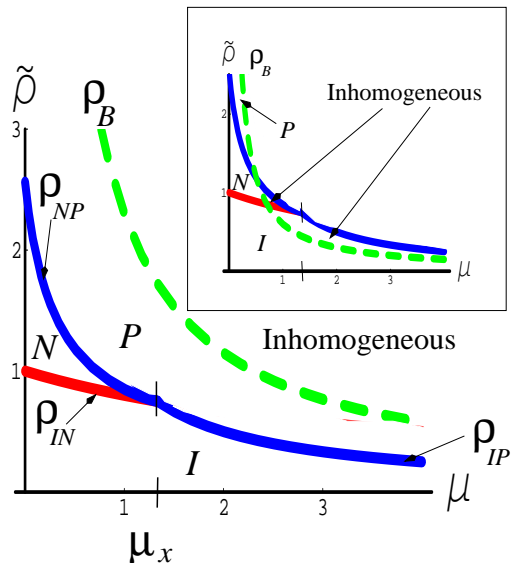}
}

\caption{The phase diagram for $g>1/4$. For $\mu>\mu_x$, where
  $\rho_{IN}$ and $\rho_{IP}$ intersect, no N state exists and the
  system goes directly from the I to the P state.  Inhomogeneous
  states form for $\rho_0>\rho_B$. The $\rho_B$ line may lie above the
  $\rho_{NP}-\rho_{IP}$ line, as shown in the main figure ($g=1$,
  $\gamma_1/\alpha=2.3$), or cross through the N and I states, as
  shown in the inset ($g=1$, $\gamma_1/\alpha=0.6$), depending on the
  value of $\gamma_1/\alpha$. (with $c_2=50$). }
\label{PD2}       
\end{figure}

II) When $g > 1/4$, the boundaries for the I-N and the N-P transitions
cross at $\mu_x=1/(g-1/4)$ where $\rho_{IN}=\rho_{IP}=\rho_{NP}$ and
the phase diagram has the topology shown in Fig.~\ref{PD2}. For
$\mu>\mu_x$
the system
goes directly from the I to the P state at $\rho_{IP}$, without an
intervening N state.  At the onset of the polarized state the
alignment tensor is again slaved to the polarization field,
\begin{math}
\rho_0
S_{ij}=\frac{b_2}{a_2}~(p_ip_j-\frac{1}{2}\delta_{ij}p^2)\;,
\end{math}
and $p_i=p_0n_i$ and  $p_0$ is determined by cubic terms
in Eq. \ref{hydrop}.

If $\gamma_1=0$, with $\gamma_0\not=0$,  the I-N transition is
independent of motor density and always occurs at $\rho_0=\rho_N$.

\paragraph{Spatially inhomogeneous states.}
Spatial inhomogeneities of course affect the stability of the
homogeneous states described above. As shown by several authors,
the rate of motor-induced filament bundling can exceed that of
filament diffusion yielding the unstable growth of density
inhomogeneities~\cite{nakazawa96,Kruse00,Kruse01,TBLMCM03}. States
with spatially varying orientational order, where the filaments
spontaneously arrange in vortex and aster structures, are also
possible~\cite{Lee01,Sankararaman04,Aranson05}. To examine the
role of spatial inhomogeneities, we have obtained coupled
equations for the first three moments of the filament
concentration defined in Eq.~(\ref{tens_expand}) by an expansion
in spatial gradients described elsewhere
\cite{TBLMCM03,TBLMCMepl}. These equations can then be used to
study the linear stability of each of the homogeneous states
against the growth of spatially-varying fluctuations in the
hydrodynamic fields. These are the fields whose characteristic
decay time are much longer than any microscopic relaxation time
and become infinitely long lived at long wavelengths. We find that
the low frequency hydrodynamic modes of this active system are
determined by fluctuations in the conserved densities and in
variables associated with broken symmetries.  A change in sign in
the decay rate of these modes signals an instability of the
macroscopic state of interest. For simplicity we only discuss here
the case of constant motor density.

\emph{Isotropic state.} This has been studied  in Ref
\cite{TBLMCM03}. The only hydrodynamic variable is the filament
density, $\rho$.  The decay rate of Fourier components of
$\delta\rho=\rho-\rho_0$ at wavevector $k$ is controlled by the
interplay of diffusion and motor-induced filament bundling
described by $\alpha$.  The homogeneous I state is unstable at
large length scales for $\rho_0>\rho_B$, with $\rho_B\sim
D_\parallel/(m_0\alpha)\sim \gamma_1/(\alpha\mu)$. The homogeneous
state is stabilized at short scales by excluded volume corrections
and higher order terms in the density gradients. The density
instability is driven entirely by the bundling rate $\alpha$,
while  $\beta$ plays no role.

\emph{Polarized state.} The hydrodynamic modes in the P state are
those associated with fluctuations in the filament density and in
the director field, ${\bf n}(\bfr,t)$, defined by ${\bf
  p}(\bfr,t)=p(\bfr,t){\bf n}(\bfr,t)$, with $|{\bf n}|=1$.
The coupled
hydrodynamic modes describing the decay of Fourier components of
density, $ \delta \rho$ and director fluctuations, $ \delta {\bf
n}={\bf n}-{\bf\hat{y}}={\bf\hat{x}}\delta n_x$ of wavevector
${\bf k}$
  are always propagating, with velocity
whose magnitude and direction are controlled by {\em both} the
activity parameters $\beta$ and $\lambda$.  For ${\bf k}$ along
the broken symmetry direction, the modes decouple (i.e., $ \delta
\rho \sim e^{z_\rho (k) t}, \delta n_x \sim e^{z_n (k) t}$) and
are given by
%
\begin{eqnarray}
&&z_\rho=ikc_1\trho\mu\tilde{\beta}
-\frac{k^2}{8}\Big[1-\frac{g\mu}{6}-20\trho\mu\tilde{\alpha}\Big]\;,\\
&&z_n=-ikc_2\trho\mu\tilde{\beta}
-\frac{5k^2}{48}\Big[1+\frac{2}{5}\trho\mu(g-6\tilde{\alpha})\Big]\;,
\end{eqnarray}
%
where $\tilde{\beta}=\beta/\gamma_1$,
$\tilde{\alpha}=\alpha/\gamma_1$, and $c_1$ and $c_2$ are numbers
of order one. We have used
$D_\parallel=\ell^2D_r/6$ and $\lambda\sim\beta$.
Like the I state, the homogeneous P state is linearly unstable for
$\rho_0>\rho_B$. The nature of the instability changes, however,
from diffusive in the I state to oscillatory in the P state,
suggesting that spatially inhomogeneous oscillatory structures,
such as vortices, may be stable at high filament or motor
densities. The rotational effects described by $\mu\sim\gamma_1$
stabilize director fluctuations, but destabilize the density.

\emph{Nematic state.} The hydrodynamic variables in the N state
are again the filament density and a director field ${\bf
n}(\bfr,t)$, defined in terms of the alignment tensor as
$S_{ij}=S_0(n_in_j-\frac{1}{2}\delta_{ij})$.
The decay of density and director fluctuations is controlled by
coupled hydrodynamic modes which are always diffusive. The modes
decouple for ${\bf k}$ along the broken symmetry direction, with
\begin{eqnarray}
&& z_\rho=-k^2\Big[\frac{1}{6}-2\trho\mu\tilde\alpha\Big]\;,\\
&&z_n= -\frac{k^2}{8}\Big[1+\frac{19}{36}\trho\mu\Big]\;.
\end{eqnarray}
Once again, the homogeneous N state is destabilized by the growth
of density fluctuations for $\rho_0>\rho_B$, while long wavelength
director fluctuations remain stable. For arbitrary direction of
${\bf k}$ relative to the direction of broken symmetry director
fluctuations also become unstable at high density, but the fastest
growing mode is always associated with the build-up of density
inhomogeneities. In contrast to the P state, the instability is
always diffusive.

To summarize, we have studied the phase behavior of polar filaments
interacting with polar clusters. We have shown that in addition to the
homogeneous \emph{isotropic} phase, both homogeneous \emph{polarized}
and \emph{nematic} states can be obtained as a function of filament
density and motor activity and polarity. The instabilities of the
homogeneous states are controlled by the bundling rate $\alpha$ and
occur for $\rho_0>\rho_B\sim \gamma_1/(\alpha\mu)$. The location of
this line in the phase diagram depends crucially on the parameter
$\gamma_1/\alpha$. If $\gamma_1/\alpha>1/g=\gamma_1/\gamma_0$, then
$\rho_B>\rho_P$ as shown in Fig.~\ref{PD2}, and the homogeneous
nematic state is always stable, when it exists.  If
$\gamma_1/\alpha<1/g=\gamma_1/\gamma_0$, then $\rho_B<\rho_P$ as shown
in the inset of Fig.~\ref{PD2}, and all homogeneous states can be
destabilized by filament bundling, albeit through different (diffusive
versus oscillatory) mechanisms.


\acknowledgments TBL acknowledges the support of the Royal
Society. MCM acknowledges support from the National Science
Foundation, grants DMR-0305407 and DMR-0219292.


\begin{thebibliography}{0}
\bibitem{meanVnote}
It was shown in Ref.~\cite{TBLMCMepl} that due to the anisotropy
of the diffusion tensor of  rods the motor cluster can induce a
net velocity of the filament pair even without external forces.
\end{thebibliography}
\end{document}